\documentstyle[prb,aps,eqsecnum,tighten,preprint]{revtex}
\begin{document}
\draft
\title{Effect of a voltage probe on the phase-coherent conductance\\ of a
ballistic chaotic cavity}
\author{P. W. Brouwer and C. W. J. Beenakker}
\address{Instituut-Lorentz, University of Leiden, P.O. Box 9506, 2300 RA
Leiden, The Netherlands}
\maketitle

\begin{abstract}
The effect of an invasive voltage probe on the phase-coherent conduction
through a ballistic chaotic cavity is investigated by random-matrix theory. The
entire distribution  $P(G)$ of the conductance $G$ is computed for the case
that the cavity is coupled to source and drain by two point contacts with a
quantized conductance of $2 e^2/h$, both in the presence ($\beta = 1$) and
absence ($\beta = 2$) of time-reversal symmetry. The loss of phase-coherence
induced by the voltage probe causes a crossover from $P(G) \propto G^{-1 +
\beta/2}$ to a Gaussian centered at $G = e^2/h$ with a $\beta$-dependent width.
\bigskip
%
\pacs{PACS numbers: 05.45.+b, 72.10.Bg, 72.15.Rn}
\end{abstract}

\section{Introduction}

A basic notion in mesoscopic physics is that the measurement of a voltage at
some point in the sample is an invasive act, which may destroy the phase
coherence throughout the whole sample. B\"uttiker introduced a simple but
realistic model for a voltage probe,\cite{Buettiker1} and used it to
investigate the transition from coherent to sequential tunneling through a
double-barrier junction, induced by the coupling to a voltage lead of the
region between the barriers. The mechanism by which the measurement of a
voltage destroys phase coherence is that electrons which enter the voltage lead
are reinjected into the system without any phase relationship. B\"uttiker's
model has been applied successfully to a variety of physical
situations,\cite{Buettiker3,BSDV,KLDV,KSEWI,Hershfield,Mello1,Mello2,BVH}
including diffusive transport in a disordered wire, ballistic transport through
quantum point contacts, and edge-channel transport in the quantum Hall effect.
In order to analyze their experimental data, Marcus et al.\cite{MWHG} proposed
to use B\"uttiker's model to describe inelastic processes in ballistic and
chaotic cavities (``quantum dots''). Here we present a detailed analysis of the
effect of a voltage probe on the entire conductance distribution of such a
system.

Several recent theoretical papers dealt with the phase-coherent conduction
through a ballistic chaotic cavity, either by means of a semiclassical
approach,\cite{BarangerJalabertStone} or by means of the supersymmetry
method,\cite{PrigodinEfetovIida,Pluhar,Mucciolo} or by random-matrix
theory.\cite{BarangerMello,JalabertPichardBeenakker,BrouwerBeenakker} Quantum
interference has a striking effect on the conductance $G$ of the quantum dot if
it is coupled to source and drain reservoirs by means of two ballistic point
contacts with a quantized conductance of $2e^2/h$. Classically, one would
expect a conductance distribution $P(G)$ which is peaked at $G = e^2/h$, since
half of the electrons injected by the source are transmitted on average to the
drain. Instead, $P(G)$ was found to
be\cite{BarangerMello,JalabertPichardBeenakker}
\begin{equation}
  P(G) \propto G^{-1 + \beta/2}, \ \ 0 \le G \le 2e^2/h,
\label{BallisticDistribution}
\end{equation}
where $\beta \in \{1,2,4\}$ is the symmetry index of the ensemble of scattering
matrices ($\beta = 1$ or $2$ in the absence or presence of a
time-reversal-symmetry breaking magnetic field; $\beta = 4$ in zero magnetic
field with strong spin-orbit scattering). Depending on $\beta$, the conductance
distribution is either uniform, peaked at zero or peaked at $2e^2/h$. As we
will show, strong coupling of the quantum dot to a voltage lead causes a
crossover from Eq.\ (\ref{BallisticDistribution}) to a Gaussian, peaked at
$e^2/h$. A small displacement of the peak of the Gaussian for $\beta=1$, and a
$\beta$-dependent width of the peak are the remnants of the weak localization
and mesoscopic fluctuation effects which are so pronounced in the case of
complete phase coherence.\cite{BarangerMello,JalabertPichardBeenakker}

A strong coupling of the voltage probe is achieved by means of a wide ballistic
lead with many scattering channels (Sec.\ \ref{sec4}). If the voltage lead
contains a single channel, we may reduce the coupling to zero by means of a
tunnel barrier in this lead (Sec.\ \ref{sec3}). Together, these two sections
cover the full range of coupling strengths. In the next section we first
formulate the problem in some more detail, and discuss the random-matrix method
used to compute the conductance distribution.

\section{Formulation of the problem}

We consider a ballistic and chaotic cavity (quantum dot) coupled by two leads
to source and drain reservoirs at voltages $V_1$ and $V_2$. A current $I = I_1
= -I_2$ is passed from source to drain via leads $1$ and $2$. A third lead is
attached to the quantum dot and connected to a third reservoir at voltage
$V_3$. This third lead is a voltage probe, which means that $V_3$ is adjusted
in such a way, that no current is drawn ($I_3 = 0$). The coupling strength of
the voltage probe is determined by the number $N$ of scattering channels
(propagating transverse modes at the Fermi-level) in lead $3$ and by the
transparency of a tunnel barrier in this lead. We assume that each of the $N$
modes has the same transmission probability $\Gamma$ through the tunnel
barrier. We restrict ourselves to the case that the current-carrying leads $1$
and $2$ are ideal (no tunnel barrier) and single-channel (a single propagating
transverse mode). This case maximizes the quantum-interference effects on the
conductance. We assume that the capacitance of the quantum dot is sufficiently
large that we may neglect the Coulomb blockade, and we will regard the
electrons to be non-interacting.

The scattering-matrix $S$ of the system has dimension $M = N + 2$ and can be
written as
\begin{equation}
  S = \left( \begin{array}{ccc} r_{11} & t_{12} & t_{13} \\
                      t_{21} & r_{22} & t_{23} \\
                      t_{31} & t_{32} & r_{33} \end{array} \right)
\end{equation}
in terms of reflection and transmission matrices $r_{ii}$ and $t_{ij}$. The
currents and voltages satisfy B\"uttiker's relations\cite{Buettiker2}
\begin{equation}
  {h \over 2 e^2} I_k = \left( N_k - R_{kk} \right) V_k - \sum_{l \neq k}
T_{kl} V_l,\ k = 1,2,3,
  \label{Buetteq}
\end{equation}
where $R_{kk} = \mbox{tr}\, r_{kk}^{\phantom \dagger} r_{kk}^{\dagger}$,
$T_{kl} = \mbox{tr}\, t_{kl}^{\phantom \dagger} t_{kl}^{\dagger}$, and $N_k$ is
the number of modes in lead $k$. The two-terminal conductance $G = I/(V_1 -
V_2)$ follows from Eq.\ (\ref{Buetteq}) with $I_1 = -I_2 = I$, $I_3 = 0$:
\begin{equation}
 G = {2e^2 \over h} \left( T_{12} + {T_{13} T_{32} \over T_{31} + T_{32}}
\right). \label{Conductance}
\end{equation}
{}From now on, we will measure $G$ in units of $2 e^2/h$.

An ensemble of quantum dots is constructed by considering small variations in
shape or Fermi energy. To compute the probability distribution $P(G)$ of the
conductance in this ensemble we need to know the distribution of the elements
of the scattering matrix. Our basic assumption, following Refs.\
\ref{BarangerMello} and \ref{JalabertPichardBeenakker}, is that for ideal leads
the scattering matrix is uniformly distributed in the space of unitary $M
\times M$ matrices. This is the circular ensemble of random-matrix
theory.\cite{Dyson,Mehta} The distribution $P_0(S)$ for the case $\Gamma = 1$
is therefore simply
\begin{equation}
  P_0(S) = {1 \over V}, \label{circ}
\end{equation}
where $V = \int d\mu$ is the volume of the matrix space with respect to the
invariant measure $d\mu$. Both $V$ and $d\mu$ depend on the symmetry index
$\beta \in \{1,2,4\}$, which specifies whether $S$ is unitary ($\beta = 2$),
unitary symmetric ($\beta = 1$), or unitary self-dual ($\beta = 4$).

A characteristic feature of the circular ensemble is that the average $\bar{S}$
of the scattering matrix vanishes. For non-ideal leads this is no longer the
case, and Eq.\ (\ref{circ}) therefore has to be modified if $\Gamma \neq 1$. In
Ref.\ \ref{BrouwerBeenakker} we showed, for a quantum dot with two non-ideal
leads, how the probability distribution $P(S)$ of the scattering matrix can be
computed by expressing the elements of the full scattering matrix $S$ (quantum
dot plus tunnel barriers) in terms of the scattering matrix $S_0$ of the
quantum dot alone (with ideal leads). A more general analysis along these
lines\cite{Brouwer} shows that for an arbitrary number of leads the
distribution takes the form of a Poisson kernel,\cite{Hua,MelloPereyraSeligman}
\begin{mathletters}
\label{Poisson}
\begin{equation}
  P(S) = c\, |\det(1 - \bar{S}^{\dagger} S)|^{-\beta M -2 + \beta},
\label{Poisson1}
\end{equation}
with normalization constant
\begin{equation}
  c = {1 \over V} [\det(1 - \bar{S}^{\dagger} \bar{S})]^{\case{1}{2}\beta M + 1
- \case{1}{2}\beta}. \label{Poisson2}
\end{equation}
In the present case of two single-channel ideal leads and one non-ideal lead
the average $\bar{S} = \int d\mu\, S P(S)$ of the scattering matrix is given by
\begin{equation}
  \bar{S}_{nm} = \left\{ \begin{array}{cl} \sqrt{1 - \Gamma} & \mbox{if $3 \le
n = m \le M$,} \\ 0 & \mbox{otherwise.} \end{array} \right.
\end{equation}
\end{mathletters}%
One verifies that for $\Gamma = 1$, $P(S)$ reduces to the distribution
(\ref{circ}) of the circular ensemble.

Eq.\ (\ref{Poisson}) holds for any $\beta \in \{1,2,4\}$. In what follows,
however, we will only consider the cases $\beta = 1,2$ of unitary or unitary
symmetric matrices, appropriate for systems without spin-orbit scattering. The
case $\beta = 4$ of unitary self-dual matrices is computationally much more
involved, and also less relevant from a physical point of view.

As indicated by B\"uttiker,\cite{Buettiker1} the cases $N = 1$ and $N > 1$ of a
single- and multi-channel voltage lead are essentially different. Current
conservation (i.e. unitarity of $S$) poses two restrictions on $T_{31}$ and
$T_{32}$: (i) $T_{31} \le 1$, $T_{32} \le 1$; and (ii) $T_{31} + T_{32} \le N$.
The second restriction is effective for $N = 1$ only. So for $N=1$, current
conservation imposes a restriction on the coupling strength of the voltage lead
to the quantum dot which is not present for $N > 1$. We treat the cases $N=1$
and $N>1$ separately, in Secs.\ \ref{sec3} and \ref{sec4}. For $N=1$ we treat
the case of arbitrary $\Gamma$, but for $N > 1$ we restrict ourselves for
simplicity to $\Gamma = 1$.

\section{Single-channel voltage lead}

\label{sec3}

In the case $N=1$, Eq.\ (\ref{Poisson}) reduces to
\begin{equation}
  P(S) = {1 \over V} \Gamma^{\beta + 1} \left(1 + (1-\Gamma)|S_{33}|^2
 -2(1-\Gamma)^{1/2}\, \mbox{Re}\, S_{33} \right)^{-\beta - 1}.
\label{PoissonKernel2}
\end{equation}
In order to calculate $P(G)$, we need to know the invariant measure $d\mu$ in
terms of a parameterization of $S$ which contains the transmission coefficients
explicitly. The matrix elements of $S$, in the case $N=1$, are related to
$R_{kk}$ and $T_{kl}$ by $S_{kk} = \sqrt{R_{kk}} e^{i \phi_{kk}}$, $S_{kl} =
\sqrt{T_{kl}} e^{i \phi_{kl}}$, where $\phi_{kl}$ are real phase shifts. When
time-reversal symmetry is broken ($\beta = 2$), we choose $R_{11}$, $R_{22}$,
$T_{12}$, $T_{21}$, $\phi_{13}$, $\phi_{23}$, $\phi_{33}$, $\phi_{32}$, and
$\phi_{31}$ as independent variables, and the other variables then follow from
unitarity of $S$. In the presence of time-reversal symmetry ($\beta = 1$), the
symmetry $S_{kl} = S_{lk}$ reduces the set of independent variables to
$R_{11}$, $R_{22}$, $T_{12}$, $\phi_{13}$, $\phi_{23}$, and $\phi_{33}$.

We compute the invariant measure $d\mu$ in the same way as in Ref.\
\ref{BarangerMello}. Denoting the independent variables in the parameterization
of $S$ by $x_i$, we consider the change $dS$ in $S$ associated with an
infinitesimal change $dx_i$ in the independent variables. The invariant
arclength $\mbox{tr}\, dS^{\dagger} dS$ defines the metric tensor $g_{ij}$
according to
\begin{equation}
  \mbox{tr}\, dS^{\dagger} dS = \sum_{i,j} g_{ij} dx_i dx_j.
\end{equation}
The determinant $\det g$ then yields the invariant measure
\begin{equation}
  d\mu = |\det g|^{1/2} \prod_{i} dx_i.
\end{equation}
The result turns out to be independent of the phases $\phi_{kl}$ and to have
the same form for $\beta = 1$ and $2$,
\begin{mathletters} \label{invmeas}
\begin{equation}
  d\mu = (\beta J)^{-1/2} \Theta(J) \prod_i dx_i.
\end{equation}
The quantity $J$ is defined by
\begin{eqnarray}
  J = \left\{ \begin{array}{l} \displaystyle 0 \
    \mbox{if $R_{11} + T_{12} > 1$ or $R_{22} + T_{21} > 1$}, \\
      4 R_{22} T_{12} T_{13} T_{23} -
      (R_{22} T_{12}  + T_{13} T_{23} - R_{11} T_{21})^2
  \ \mbox{otherwise}, \end{array} \right.
\end{eqnarray}
\end{mathletters}%
and $\Theta(J) = 1$ if $J > 0$ and $\Theta(J) = 0$ if $J \le 0$. The
independent variables $x_i$ are different, however, for $\beta = 1$ and $\beta
= 2$ --- as indicated above.

We have calculated the probability distribution of the conductance from Eqs.\
(\ref{Conductance}), (\ref{PoissonKernel2}), and (\ref{invmeas}). The results
are shown in Fig.\ \ref{fig1}, for several values of $\Gamma$. For $\Gamma = 0$
(uncoupled voltage lead), $P(G)$ is given
by\cite{BarangerMello,JalabertPichardBeenakker}
\begin{equation}
  P(G) = \left\{ \begin{array}{ll}
   \case{1}{2} G^{-1/2} & \mbox{if $\beta$ = 1}, \\
   1 & \mbox{if $\beta$ = 2}. \end{array} \right.
\end{equation}
For $\Gamma=1$ (maximally coupled single-channel voltage lead), we find
\begin{equation}
 P(G) =
    \left\{ \begin{array}{lcl}
    \displaystyle 2 - 2G
    && \displaystyle \mbox{if $\beta = 1$}, \\
    \displaystyle \case{4}{3}
    \left[2 G- 2G^2 - (3 G^2 - 2 G^3)\ln G\ -
    (1 - 3 G^2 + 2 G^3) \ln(1 - G) \right] &&
    \displaystyle \mbox{if $\beta = 2$}. \end{array} \right.
\end{equation}

The average $\langle G \rangle$ and variance $\mbox{var}\, G$ of the
conductance can be calculated in closed form for all $\Gamma$. We find that
$\langle G \rangle$ is independent of $\Gamma$,
\begin{equation}
 \langle G \rangle =
    \left\{ \begin{array}{lcl}
    \displaystyle \case{1}{3}
    && \displaystyle \mbox{if $\beta = 1$}, \\
    \displaystyle  \case{1}{2}
    && \displaystyle \mbox{if $\beta = 2$}. \end{array} \right.
\end{equation}
The variance does depend on $\Gamma$,
\begin{equation}
 \mbox{var}\, G =
    \left\{ \begin{array}{lcl}
    \displaystyle \case{1}{45} \left(1 - \Gamma \right)^{-2}
    \left({4 - 11 \Gamma + 7 \Gamma^2 - 3 \Gamma^2 \ln \Gamma}\right)
    && \displaystyle \mbox{if $\beta = 1$}, \\
    \displaystyle \case{1}{36} {\left( 1 - \Gamma \right) }^{-3}
  {\left({3 - 11\,\Gamma + 17\,{\Gamma^2} - 9\,{\Gamma^3} + 4\,{\Gamma^3}\,\ln
     \Gamma} \right)}
    && \displaystyle \mbox{if $\beta = 2$}. \end{array} \right.
\end{equation}
The breaking of phase coherence caused by a single-channel voltage lead is not
strong enough to have any effect on the average conductance, which for $\beta =
1$ remains below the classical value of $\case{1}{2}$. The variance of the
conductance is reduced somewhat when $\Gamma$ is increased from $0$ to $1$, but
remains finite. (For $\beta = 1$ the reduction is with a factor $\case{5}{8}$,
for $\beta = 2$ with a factor $\case{5}{9}$.) We will see in the next section,
that the complete suppression of quantum interference effects requires a
voltage lead with $N \gg 1$. Then $\langle G \rangle \rightarrow \case{1}{2}$
and $\mbox{var}\, G \rightarrow 0$.

\section{Multi-channel voltage lead}

\label{sec4}

Now we turn to the case of a multi-channel ideal voltage lead ($N > 1$, $\Gamma
= 1$). Current conservation yields:
\begin{equation}
  \begin{array}{lclcl}
  T_{13} &=& 1 - R_{11} - T_{12} &=& 1 - |S_{11}|^2 - |S_{12}|^2, \\
  T_{31} &=& 1 - R_{11} - T_{21} &=& 1 - |S_{11}|^2 - |S_{21}|^2, \\
  T_{32} &=& 1 - R_{22} - T_{12} &=& 1 - |S_{12}|^2 - |S_{22}|^2.
  \end{array}
\end{equation}
To determine $P(G)$ it is thus sufficient to know the distribution
$\tilde{P}(S_{11},S_{12},S_{21},S_{22})$ of the matrix elements $S_{kl}$ with
$k,l \le 2$. This marginal probability distribution has been calculated by
Mello and coworkers\cite{PereyraMello} for arbitrary dimension $M \ge 4$ of
$S$. As in Sec.\ \ref{sec3} we parameterize $S_{kl} = \sqrt{T_{kl}} e^{i
\phi_{kl}}$ if $k \neq l$ and $S_{kk} = \sqrt{R_{kk}} e^{i \phi_{kk}}$ ($k, l
\le 2$). We abbreviate $\prod_i dy_i \equiv dR_{11} dR_{22} dT_{12} dT_{22}
\prod_{k,l=1}^{2} d\phi_{kl}$. For the cases $\beta = 1,2$ one then
has\cite{PereyraMello}
\begin{mathletters} \label{multidensity}
\begin{equation}
  d \tilde{P} = \left\{ \begin{array}{lcl}
  \displaystyle c_1
  \delta(T_{12} - T_{21}) \delta(\phi_{12} - \phi_{21}) F^{(M - 5)/2}
  \Theta(F) \prod_i dy_i && \mbox{if $\beta = 1$},
  \label{FiniteDistribution1} \\
  \displaystyle c_2 F^{M-4} \Theta(F) \prod_i dy_i && \mbox{if $\beta = 2$},
  \label{FiniteDistribution2}
  \end{array} \right.
\end{equation}
where $F$ is defined by
\begin{eqnarray}
  F &=& \left\{ \begin{array}{l} 0 \ \mbox{if $R_{11} + T_{12} > 1$ or $R_{22}
+ T_{21} > 1$,} \\
  (1-R_{11})(1-R_{22}) + (1-T_{12})(1-T_{21}) - 1 \\ \hspace{1cm} -\, 2 (R_{11}
R_{22} T_{12} T_{21})^{1/2} \cos(\phi_{11} + \phi_{22} - \phi_{12} - \phi_{21})
\  \mbox{otherwise}. \end{array} \right.
\end{eqnarray}
\end{mathletters}%
The coefficients $c_1$ and $c_2$ are normalization constants. Calculation of
the probability distribution of the conductance is now a matter of quadrature.

Results are shown in Fig.\ \ref{fig2}, for $N$ up to $10$. As $N$ increases,
$P(G)$ becomes more and more sharply peaked around $G = \case{1}{2}$. In the
limit $N \rightarrow \infty$, $P(G)$ approaches a Gaussian, with mean and
variance given by
\begin{eqnarray}
  \langle G \rangle &=& \left\{ \begin{array}{lcl} \displaystyle
     \case{1}{2} - \case{1}{2} N^{-1} + {\cal O}(N^{-2}) &&
     \mbox{if $\beta = 1$}, \\
     \displaystyle \lefteqn{\case{1}{2}}
     \hphantom{\case{3}{4} N^{-2} + {\cal O}(N^{-3})} &&
     \mbox{if $\beta = 2$}, \end{array} \right. \\
 \mbox{var}\, G &=& \left\{ \begin{array}{lcl}
     \lefteqn{\case{3}{4} N^{-2} + {\cal O}(N^{-3})}
     \hphantom{\displaystyle
       \case{1}{2} - \case{1}{2} N^{-1} + {\cal O}(N^{-2})} &&
     \mbox{if $\beta = 1$}, \\
     \displaystyle \case{1}{4} N^{-2} + {\cal O}(N^{-3}) &&
     \mbox{if $\beta = 2$}. \end{array} \right.
\end{eqnarray}
The variance of $G$ is reduced by a factor $3$ when time-reversal symmetry is
broken in the limit $N \rightarrow \infty$. The offset of $\langle G \rangle$
from $\case{1}{2}$ when $\beta = 1$ is a remnant of the weak localization
effect.

\section{Conclusion}

We have calculated the entire probability distribution of the conductance of a
quantum dot in the presence of a voltage probe, for single-channel point
contacts to source and drain, in the presence and absence of time-reversal
symmetry (no spin-orbit scattering). The average conductance is not changed if
a single-channel voltage lead containing a tunnel barrier is attached, but the
shape of the distribution changes considerably. A strikingly simple result is
obtained for a single-channel ballistic voltage lead in zero magnetic field
($N=1$, $\Gamma=1$, $\beta=1$), when $P(G) = 2 - 2G$, to be compared with $P(G)
= \case{1}{2} G^{-1/2}$ without the voltage
probe.\cite{BarangerMello,JalabertPichardBeenakker} (In both cases $G \in
[0,1]$ is measured in units of $2e^2/h$.) When the number $N$ of channels in
the voltage lead is increased, the probability distribution becomes sharply
peaked around $G = \case{1}{2}$. Both the width of the peak and the deviation
of its center from $\case{1}{2}$ scale as $1/N$ for $N \gg 1$. The width is
reduced by a factor $\sqrt{3}$ upon breaking the time-reversal symmetry.

The loss of phase coherence induced by a voltage probe can be investigated
experimentally by fabricating a cavity with three leads attached to it.
Furthermore, as emphasized by Marcus et al.,\cite{MWHG} the inelastic
scattering which occurs at finite temperatures in a quantum dot might well be
modeled effectively by an imaginary voltage lead.

This research was supported by the ``Ne\-der\-land\-se or\-ga\-ni\-sa\-tie voor
We\-ten\-schap\-pe\-lijk On\-der\-zoek'' (NWO) and by the ``Stich\-ting voor
Fun\-da\-men\-teel On\-der\-zoek der Ma\-te\-rie'' (FOM).

\begin{figure}
\caption{Distribution of the conductance $G$ (in units of $2 e^2/h$) for a
single-channel voltage lead ($N=1$). The voltage lead contains a tunnel barrier
with transmission probability $\Gamma$, which varies from $0$ to $1$ with
increments of $0.2$. (a): time-reversal symmetry ($\beta = 1$); (b): broken
time-reversal symmetry ($\beta = 2$). The quantum dot is shown schematically in
the inset.\label{fig1}}
\end{figure}

\begin{figure}
\caption{Conductance distribution for a multi-channel ideal voltage lead
($\Gamma = 1$). The number $N$ of transverse modes in the lead varies from $1$
to $10$ with increments of $1$ (solid curves). The dotted curve is the
distribution in the absence of a voltage lead. The cases $\beta = 1$ and $2$
are shown in (a) and (b) respectively.\label{fig2}}
\end{figure}

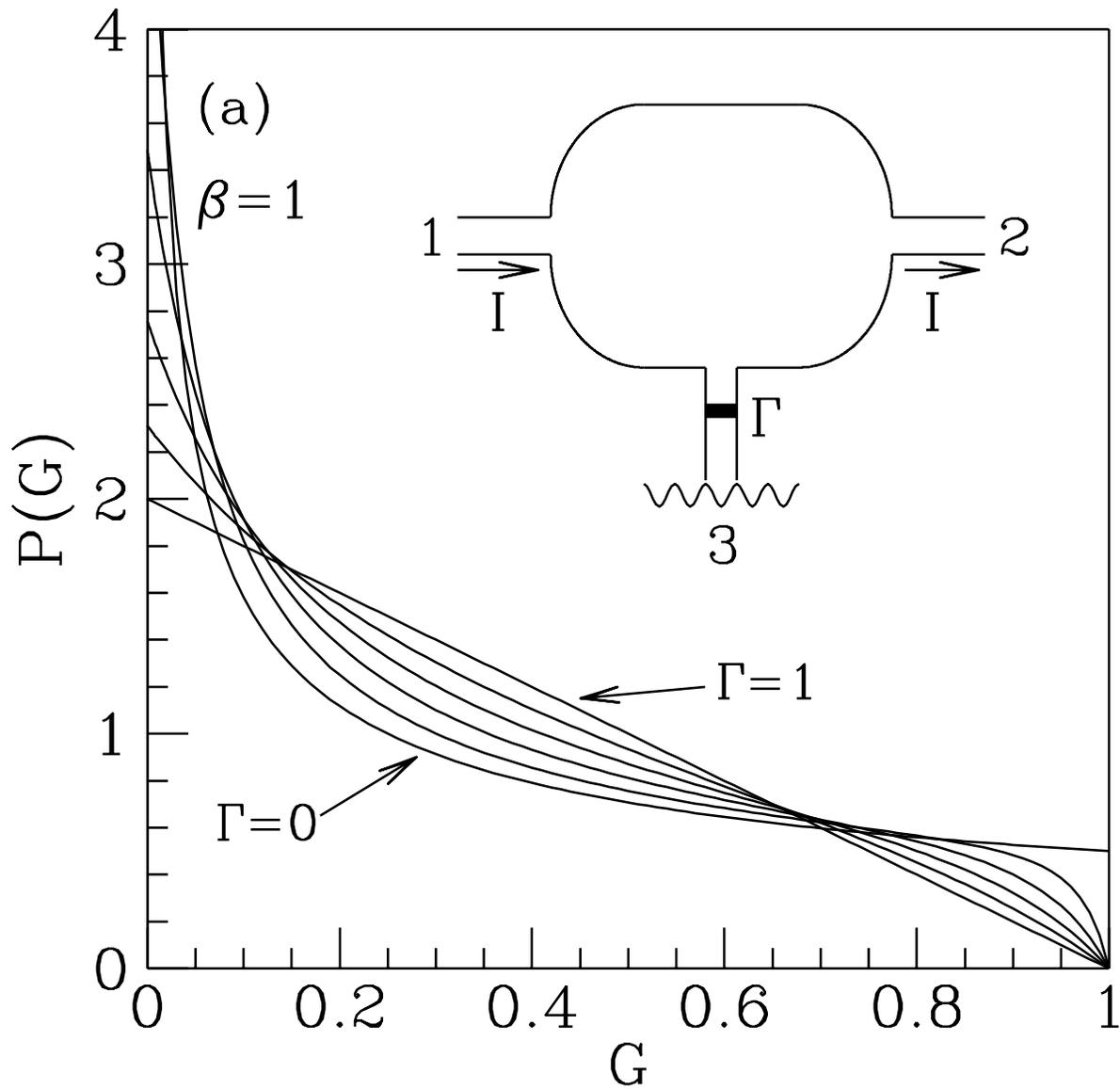
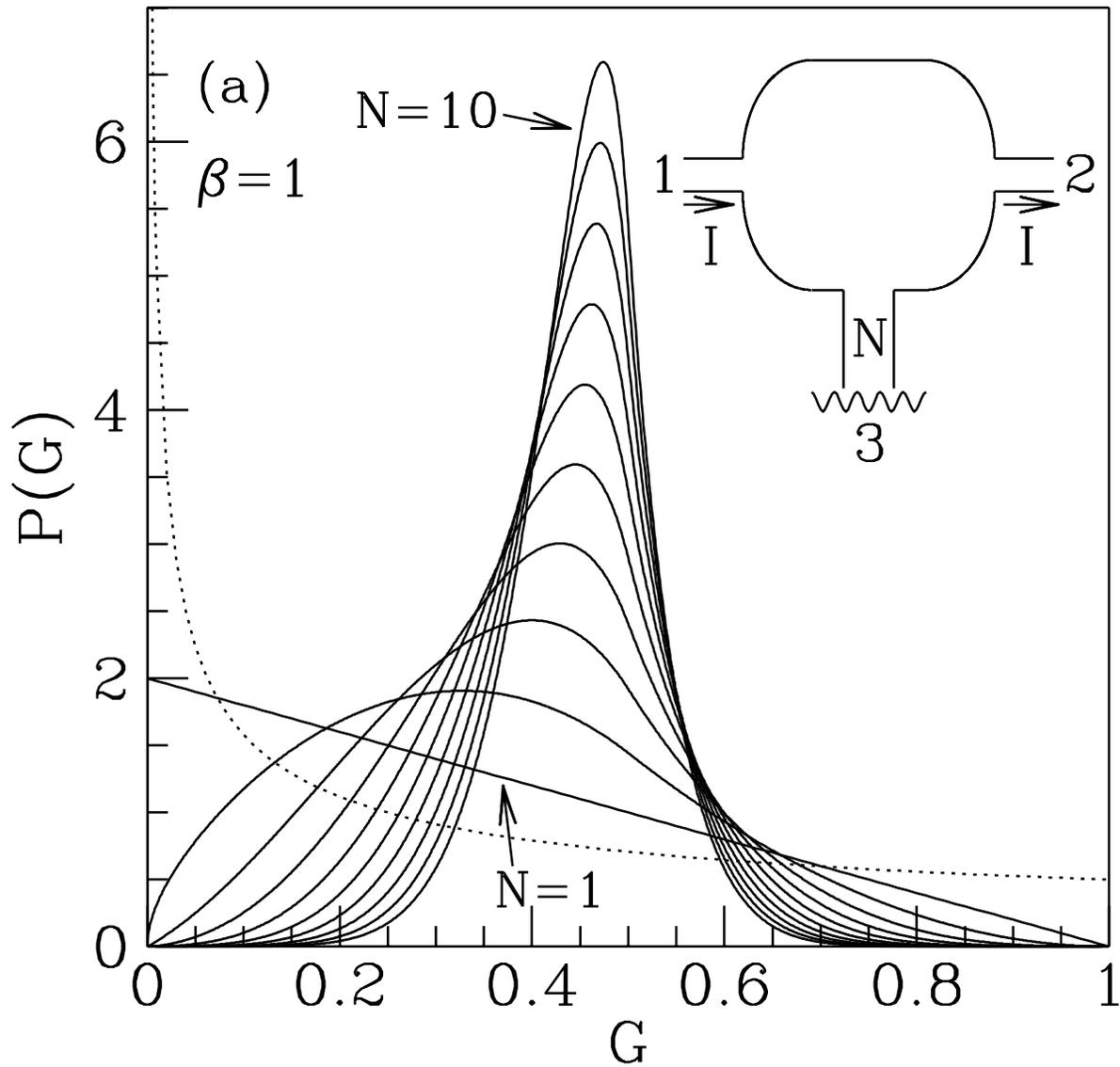
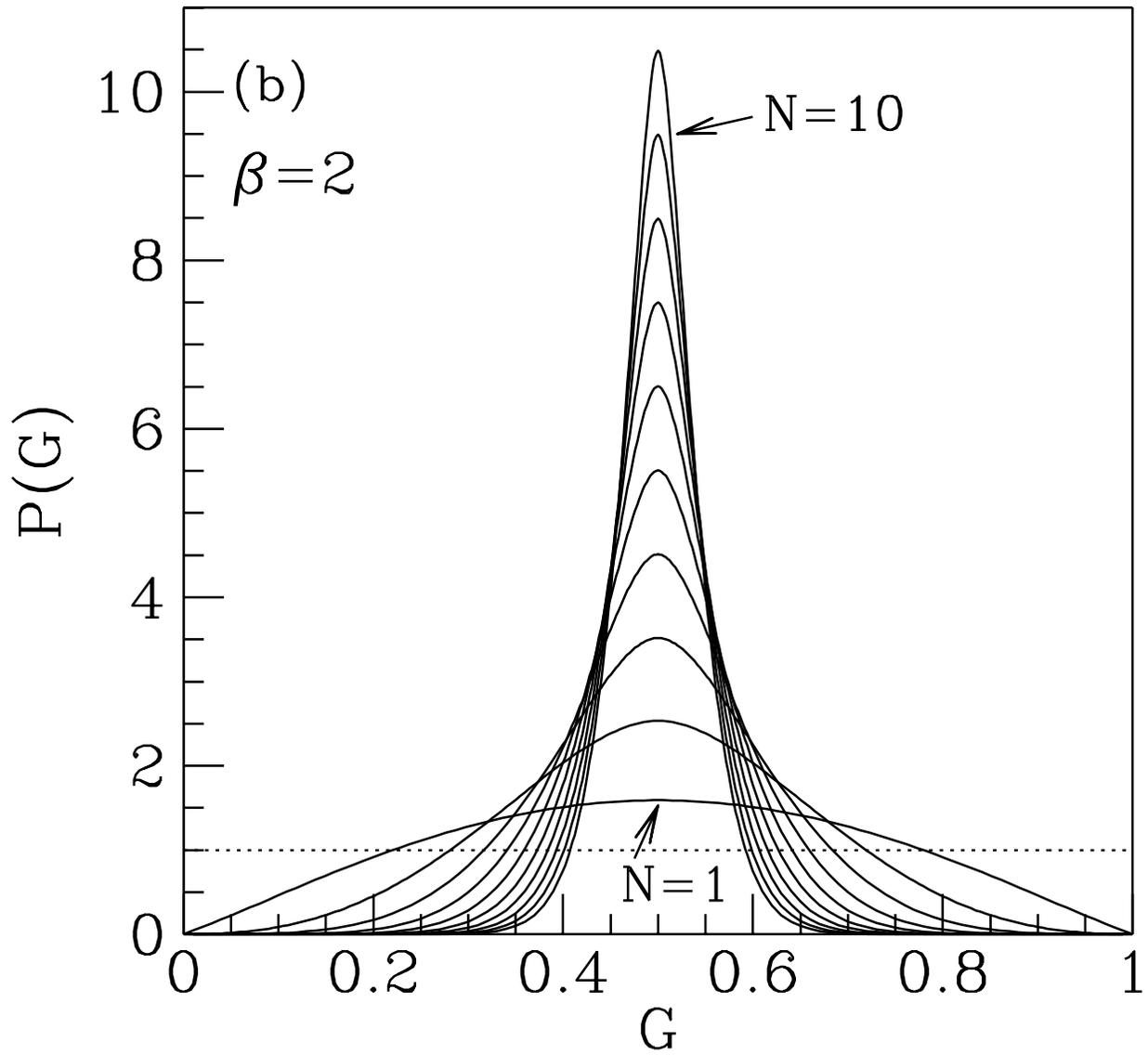
\end{document}